\documentclass[11pt,twoside]{article}
\input{epsf.sty}
\usepackage{mathrsfs}
\usepackage{amsmath}
\usepackage{amssymb}
\usepackage{graphicx}
\usepackage{subfigure}
\title{ Integrable system with peakon, weak kink, and kink-peakon interactional solutions}
\author{Baoqiang Xia$^{1}$\footnote{E-mail address:
xiabaoqiang@126.com}, ~~Zhijun Qiao$^{2}$\footnote{Corresponding author. E-mail address:
qiao@utpa.edu}, ~~Jibin Li$^{3}$\footnote{E-mail address:
lijb@zjnu.cn},
\\
$^1$School of Mathematics and Statistics, Jiangsu Normal
University,\\
 Xuzhou, Jiangsu 221116, P. R. China
\\ $^{2}$Department of Mathematics, University of Texas-Pan American, \\Edinburg, Texas 78541, USA
\\
 $^3$Department of Mathematics,
 Zhejiang Normal University,\\  Jinhua, Zhejiang 321004, P. R. China}

\date{}
\begin{document}
\maketitle
\begin{abstract}
In this paper, we study an integrable system with both quadratic and cubic nonlinearity: $m_t=bu_x+\frac{1}{2}k_1\left[m(u^2-u^2_x)\right]_x+\frac{1}{2}k_2(2m u_x+m_xu)$, $m=u-u_{xx}$,
where $b$, $k_1$ and $k_2$ are arbitrary constants. This model is kind of a cubic generalization of the Camassa-Holm (CH) equation: $m_t+m_xu+2mu_x=0$.  The equation is shown integrable with its Lax pair, bi-Hamiltonian structure, and infinitely many conservation laws. In the case $b=0$, the peaked soliton (peakon) and multi-peakon solutions are studied. In particular, the two-peakon dynamical system is explicitly presented and their collisions are investigated in details. In the case $b\neq0$, the weak kink and kink-peakon interactional solutions are found. Significant difference from the CH equation is analyzed through a comparison.
In the paper, we also study all possible smooth one-soliton solutions for the system.


\end{abstract}
\newpage

\section{ Introduction}
The Camassa-Holm (CH) equation
\begin{eqnarray}
m_t-bu_x+2m u_x+m_xu=0, \quad m=u-u_{xx},
\label{bCH}
\end{eqnarray}
was derived by Camassa and Holm \cite{CH} as a shallow water
wave model.  In recent years, this equation has attracted much attention in soliton theory.  In the literature,
this equation was implicitly implied in the paper of Fuchssteiner and Fokas on hereditary symmetries as a very special case \cite{FF1}.
However, since the work of Camassa and Holm \cite{CH}, various studies on this equation have been remarkably developed. 
The CH equation possesses many important integrable properties. 
For instance, it admits Lax representation, bi-Hamiltonian structures, multi-soliton solutions, and algebro-geometric solutions \cite{CH2}-\cite{LZ}. 
Also, it is integrable by the inverse scattering transformation \cite{C1,C2}. 
The most remarkable feature of the CH equation (\ref{bCH}) is that it admits peaked
soliton (peakon) solutions in the case $b=0$ \cite{CH,CH2}. A peakon is a weak
solution in some Sobolev space with corner at its crest.
The stability and interaction of peakons were discussed in several references \cite{CS1}-\cite{JR}.
In addition to the CH equation being an integrable model with peakon solutions, other integrable peakon models have been found. Those models include the Degasperis-Procesi equation \cite{DP1}-\cite{DP3} and the cubic nonlinear peakon equations \cite{Q1}-\cite{NV1}.

The present paper focuses on the following equation with both quadratic and cubic nonlinearity:
\begin{eqnarray}
m_t=bu_x+\frac{1}{2}k_1\left[m(u^2-u^2_x)\right]_x+\frac{1}{2}k_2(2m u_x+m_xu), \quad m=u-u_{xx},\label{bCHFFORQ}
\end{eqnarray}
where $b$, $k_1$ and $k_2$ are three arbitrary constants. It is clear that  equation (\ref{bCHFFORQ}) is reduced to the CH equation (\ref{bCH}) when we take $k_1=0$, $k_2=-2$.  For $k_1=-2$, $k_2=0$, equation (\ref{bCHFFORQ}) is exactly the cubic nonlinear equation:
\begin{eqnarray}
m_t-bu_x+\left[m(u^2-u^2_x)\right]_x=0, \quad m=u-u_{xx},\label{qe}
\end{eqnarray}
which was derived independently by Fokas \cite{Fo}, by Fuchssteiner \cite{Fu}, by Olver and Rosenau \cite{OR}, by Qiao \cite{Q1,Q2}, and by Novikov \cite{NV1}, based on different points of view.
Recently, the non-periodic peakon solution of equation (\ref{qe}) in the case $b=0$  was derived in \cite{GLOQ,HM1,HM2}, and the periodic peakon solution was derived in \cite{HM1}.

Equation (\ref{bCHFFORQ}) is actually
a linear combination of CH equation (\ref{bCH}) and cubic nonlinear equation (\ref{qe}).
Therefore, we may view equation (\ref{bCHFFORQ}) as a generalization of the CH equation, or simply call equation (\ref{bCHFFORQ}) a generalized CH equation.
This structure is very similar to the one in dealing with the Gardner equation, known as a linear combination of KdV and mKdV equations, which has important applications in various areas of physics \cite{GE1,GE2}.
In fact, equation (\ref{bCHFFORQ}) may be derived by applying tri-Hamiltonian duality \cite{OR} to the bi-Hamiltonian Gardner equation. We also notice that by some appropriate rescaling, equation (\ref{bCHFFORQ}) was implied in the papers of Fokas and Fuchssteiner \cite{Fo,Fu}, where it was derived from the two-dimensional hydrodynamical equations and it models the unidirectional propagation of the fluid lying above a horizontal flat bottom.

In this paper, we first present the Lax representation, bi-Hamiltonian structure and infinitely many conservation laws for equation (\ref{bCHFFORQ}). This indicates that equation (\ref{bCHFFORQ}) is completely integrable. Then we show that this equation possesses the single peakon of traveling wave type as well as multi-peakon solutions in the case $b=0$. In particular, for the one-peakon case, we find that the complex peakon (i.e. pekaon with complex coefficient) appears by properly choosing the parameters $k_1$ and $k_2$. For the two-peakon case, the peakon dynamical system is explicitly presented and their collisions are discussed in details. Significant difference between the CH equation (\ref{bCH}) and the generalized CH equation (\ref{bCHFFORQ}) (in the case $b=0$) is analyzed through a comparison.

Another purpose of this paper is to show that equation (\ref{bCHFFORQ}) with $k_2=0$ (namely, cubic nonlinear equation (\ref{qe})) allows the weak kink solution in the case $b\neq0$. Different from the multi-peakon solutions in the form of linear superpositions of the single-peakon, equation (\ref{bCHFFORQ}) with $k_2=0$ and $b\neq 0$ does not allow the multi-kink solution in the form of the superpositions of single-kink. However, we find that equation (\ref{bCHFFORQ}) with $k_2=0$ and $b\neq 0$ allows the solutions in the form of the superpositions of single-kink and multi-peakon. In particular, the weak kink and kink-peakon interactional solutions are shown and plotted. Within our knowledge, this is probably the first time discussing the weak kink and kink-peakon interactional solutions. In the paper, we also study all possible smooth one-soliton solutions.

\section{Lax pair, bi-Hamiltonian structure and conservation laws}

Let us consider the following pair of linear spectral problems
\begin{eqnarray}
\left(\begin{array}{c}\phi_{1}\\\phi_{2} \end{array}\right)_x&=&
U\left(\begin{array}{c} \phi_{1}\\\phi_{2}
\end{array}\right),\quad
U=\frac{1}{2}\left( \begin{array}{cc} -\alpha & \lambda m\\
 -k_1\lambda m-k_2\lambda &  \alpha \\ \end{array} \right),
\label{cCFLPs}\\
\left(\begin{array}{c}\phi_{1}\\\phi_{2} \end{array}\right)_t&=&
V\left(\begin{array}{c} \phi_{1}\\\phi_{2}
\end{array}\right),\quad V=-\frac{1}{2}\left( \begin{array}{cc} A & B \\ C & -A \\ \end{array} \right),
\label{cCFLPt}
\end{eqnarray}
where $\alpha=\sqrt{1-\lambda^2b}$, $b$ is an arbitrary constant, $\lambda$ is a spectral parameter, and
\begin{eqnarray}
\begin{split}
A=& \lambda^{-2}\alpha+\frac{\alpha}{2}k_1(u^2-u_x^2)+\frac{1}{2}k_2(\alpha u-u_x),
\\ B=& -\lambda^{-1}(u-\alpha u_x)-\frac{1}{2}\lambda m\left[ k_1(u^2-u_x^2)+k_2 u\right],
\\ C=&\lambda^{-1}\left[k_1(u+\alpha u_x)+k_2\right]+\frac{1}{2}\lambda \left[k_1^2m(u^2-u_x^2)+k_1k_2(m u+u^2-u_x^2)+k_2^2u\right].
\end{split}
\end{eqnarray}
One can easily see that the compatibility condition of (\ref{cCFLPs}) and (\ref{cCFLPt}) is
\begin{eqnarray}
U_t-V_x+[U,V]=0.\label{cc}
\end{eqnarray}
Substituting the expressions of $U$ and $V$ into (\ref{cc}), we find that (\ref{cc}) is nothing
but equation (\ref{bCHFFORQ}), namely, (\ref{cCFLPs}) and (\ref{cCFLPt}) are the Lax pair of equation (\ref{bCHFFORQ}).

A direct computation shows that equation (\ref{bCHFFORQ}) has the following bi-Hamiltonian structure
\begin{eqnarray}
m_t=J \frac{\delta H_1}{\delta m}=K \frac{\delta H_2}{\delta m},\label{HC}
\end{eqnarray}
where
\begin{eqnarray}
J&=& k_1\partial m\partial^{-1} m\partial+\frac{1}{2}k_2(\partial m+m\partial)+b\partial,
\quad H_1=\frac{1}{2}\int_{-\infty}^{+\infty}(u^2+u_x^2)dx,
\\ K&=&\partial-\partial^3,
\quad H_2=\frac{1}{8}\int_{-\infty}^{+\infty}(k_1u^4+2k_1u^2u_x^2-\frac{1}{3}k_1u_x^4+2k_2u^3+2k_2uu_x^2+4bu^2)dx.
\label{JC}
\end{eqnarray}

Let us now construct conservation laws of equation (\ref{bCHFFORQ}). Let $\omega=\frac{\phi_2}{\phi_1}$, then $\omega$ satisfies the following Riccati equation
\begin{eqnarray}
\omega_x=-\frac{1}{2}\lambda(k_1 m+k_2)+\alpha \omega-\frac{1}{2}\lambda m \omega^2.
\label{ric}
\end{eqnarray}
Based on (\ref{cCFLPs}) and (\ref{cCFLPt}), we obtain
\begin{eqnarray}
(\ln \phi_1)_x=-\frac{\alpha}{2}+\frac{1}{2}\lambda m\omega, \quad (\ln \phi_1)_t=-\frac{1}{2}A-\frac{1}{2}B\omega,
\label{lnp}
\end{eqnarray}
which yields the conservation law of equation (\ref{bCHFFORQ}):
\begin{eqnarray}
\rho_t=F_x,
\label{CL}
\end{eqnarray}
where
\begin{eqnarray}
\begin{split}
\rho&=m\omega,
\\F&=\frac{1}{2} m\left[ k_1(u^2-u_x^2)+k_2 u\right]\omega-\frac{1}{2}\left[\alpha k_1(u^2-u_x^2)+k_2(\alpha u-u_x)\right]\lambda^{-1}+(u-\alpha u_x)\omega\lambda^{-2}.
\end{split}
\label{rj}
\end{eqnarray}
Usually $\rho$ and $F$ are called a conserved density and an associated flux,
respectively. Next, we derive the explicit forms of conservation densities in the case $b=0$. In this case, equation (\ref{ric}) becomes
\begin{eqnarray}
\omega_x=-\frac{1}{2}\lambda(k_1 m+k_2)+ \omega-\frac{1}{2}\lambda m \omega^2.
\label{ric2}
\end{eqnarray}
We expand $\omega$ in terms of negative powers of $\lambda$ as below:
\begin{equation}
\omega=\sum_{j=0}^{\infty}\omega_j\lambda^{-j}.\label{oe1}
\end{equation}
Substituting (\ref{oe1}) into (\ref{ric2}) and equating the coefficients of powers of $\lambda$, we arrive at
\begin{eqnarray}
\begin{split}
\omega_{0}&=\frac{1}{m}\sqrt{-k_1m^2-k_2m}, \quad \omega_{1}=\frac{2k_1 m^{2}+k_2(2m+m_x)}{2k_1m^3+2k_2m^2},
\\
\omega_{j+1}&=\frac{1}{m\omega_0}\left[\omega_j-\omega_{j,x}-\frac{1}{2}m\sum_{i+k=j+1,~i,k\geq 1}\omega_i\omega_k\right],\quad j\geq 1.
\end{split}
\label{wj}
\end{eqnarray}
Inserting (\ref{oe1}) and (\ref{wj}) into (\ref{rj}), we finally get the following infinitely many conserved densities and the associated fluxes
\begin{eqnarray}
\begin{split}
\rho_{0}&=\sqrt{-k_1m^2-k_2m}, ~ F_0=\frac{1}{2}\left[k_1(u^2-u_x^2)+k_2u\right]\sqrt{-k_1m^2-k_2m},
\\
\rho_{1}&=\frac{2k_1 m^{2}+k_2(2m+m_x)}{2k_1m^3+2k_2m^2}, ~ F_1=-\frac{1}{2}\left[k_1(u^2-u_x^2)+k_2(u-u_x)\right]+\frac{1}{2}\left[k_1(u^2-u_x^2)+k_2u\right]\rho_1,
\\
\rho_{j+1}&=\frac{1}{\omega_0}\left(\omega_j-\omega_{j,x}-\frac{1}{2}m\sum_{i+k=j+1,~i,k\geq 1}\omega_i\omega_k\right),\quad j\geq 1,
\\
F_{j+1}&=(u-u_x)\omega_{j-1}+\frac{1}{2}\left[k_1(u^2-u_x^2)+k_2u\right]\rho_{j+1},\quad j\geq 1.
\end{split}
\label{rjj}
\end{eqnarray}

We can also consider the expansions of $\omega$ in the positive powers of $\lambda$:
\begin{equation}
\omega=\sum_{j=0}^{\infty}\omega_j\lambda^{j}.\label{oe2}
\end{equation}
Substituting (\ref{oe2}) into equation (\ref{ric2}) and comparing powers of $\lambda$ lead to
\begin{eqnarray}
\begin{split}
\omega_{2j}&=0, \quad j\geq 0,
\\
\omega_{1}&=\frac{1}{2}[k_1(u+u_x)+k_2] , \quad \omega_{2j+1,x}-\omega_{2j+1}=-\frac{1}{2}m\sum_{i+k=2j,~i,k\geq 0}\omega_i\omega_k,\quad j\geq 1.
\end{split}
\label{wj2}
\end{eqnarray}
Equation (\ref{wj2}) shows that one may solve a first-order differential equation to obtain nontrivial $\omega_{2j+1}$, $j\geq 1$ in this
sequence. For brevity, we omit the result since this may involve in nonlocal expressions in $u$.

\section{Peakon solutions in the case $b=0$}

Applying the operator $(1-\partial_x^2)^{-1}$ to equation (\ref{bCHFFORQ}), we obtain
\begin{eqnarray}
\begin{split}
u_t=&\frac{1}{6}k_1(3u^2u_x-u_x^3)+\frac{1}{2}k_2uu_x+\frac{1}{6}k_1(1-\partial_x^2)^{-1}u_x^3
\\&+\frac{1}{2}\partial_x(1-\partial_x^2)^{-1}\left(2bu+k_1(uu_x^2+\frac{2}{3}u^3)+k_2(u^2+\frac{1}{2}u_x^2)\right).
\end{split}\label{cwf}
\end{eqnarray}
Taking the convolution with the Green's function $G(x)=\frac{1}{2}\exp(-|x|)$ for the Helmholtz operator $(1-\partial_x^2)$, equation (\ref{cwf}) can be rewritten as
\begin{eqnarray}
\begin{split}
u_t=&\frac{1}{6}k_1(3u^2u_x-u_x^3)+\frac{1}{2}k_2uu_x+\frac{1}{6}k_1G(x)*u_x^3
\\&+\frac{1}{2}\partial_x\left(G(x)*[2bu+k_1(uu_x^2+\frac{2}{3}u^3)+k_2(u^2+\frac{1}{2}u_x^2)]\right).
\end{split}\label{cwf2}
\end{eqnarray}
In this section, we derive the single and multi-peakon solutions of equation (\ref{bCHFFORQ}) in the case $b=0$.

\subsection{Single-peakon solutions}
Let us suppose the single-peakon solution of equation (\ref{bCHFFORQ}) with $b=0$ in the form of
\begin{eqnarray}
u=Ce^{-\mid x-ct\mid},\label{ocp}
\end{eqnarray}
where $C$ is to be determined. The derivatives of expression (\ref{ocp}) do not exist at $x=ct$, thus (\ref{ocp}) can not satisfy equation (\ref{bCHFFORQ}) with $b=0$ in the classical sense. However, in the weak sense, we can write out the expressions of $u_x$, $u_t$ and $m$ with help of distribution:
\begin{eqnarray}
u_x=-Csgn(x-ct)e^{-\mid x-ct\mid}, \quad u_t=cCsgn(x-ct)e^{-\mid x-ct\mid}, \quad m=2C\delta(x-ct). \label{ocpd}
\end{eqnarray}
Substituting (\ref{ocp}) and (\ref{ocpd}) into the weak form (\ref{cwf2}) with $b=0$, we are able to find that $C$ should satisfy
\begin{eqnarray}
\frac{1}{3}k_1C^2+\frac{1}{2}k_2C+c=0. \label{C1}
\end{eqnarray}
For $k_1=0$, $k_2\neq 0$, we obtain $C=\frac{-2c}{k_2}$. In particular, when $k_2=-2$, (\ref{ocp}) is exactly the single-peakon solution $u=ce^{-\mid x-ct\mid}$ of the CH equation (\ref{bCH}) with $b=0$. For $k_1=-2$ and $k_2=0$, we recover the single-peakon solution $u=\pm \sqrt{\frac{3c}{2}}e^{-\mid x-ct\mid}$ of the cubic nonlinear CH equation (\ref{qe}) with $b=0$ \cite{GLOQ,HM1,HM2}. In general, for $k_1\neq 0$, we may obtain
\begin{eqnarray}
C=\frac{-3\left(\sqrt{3}k_2\pm \sqrt{3k_2^2-16k_1c}\right)}{4\sqrt{3}k_1}. \label{sC1}
\end{eqnarray}
If $3k_2^2-16k_1c\geq0$, then $C$ is a real number. For example, we choose $k_1=k_2=-2$ and $c>0$, then the corresponding single-peakon solution is $u=\frac{-3\pm\sqrt{9+24c}}{4}e^{-\mid x-ct\mid}$. If $3k_2^2-16k_1c<0$, then $C$ is a complex number. This means the peakon solution with complex coefficient is obtained. For example, we set $k_2=-2$ and $k_1=\frac{1}{c}$, then we have $u=\frac{(3\pm\sqrt{3}i)c}{2}e^{-\mid x-ct\mid}$.

\subsection{Two-peakon solutions and their dynamics}
Let us assume the two-peakon solution to equation (\ref{bCHFFORQ}) with $b=0$ has the following form
\begin{eqnarray}
u=p_1(t)e^{-\mid x-q_1(t)\mid}+p_2(t)e^{-\mid x-q_2(t)\mid}.\label{tcp}
\end{eqnarray}
The expression of $u$ has two peaks (or troughs) at positions $x=q_1(t)$ and $x=q_2(t)$. $u_x$, $u_t$ and $m$ have the following weak forms:
\begin{eqnarray}
\begin{split}
u_x&=-p_1sgn(x-q_1)e^{-\mid x-q_1\mid}-p_2sgn(x-q_2)e^{-\mid x-q_2\mid},
\\ u_t&=p_{1,t}e^{-\mid x-q_1\mid}+p_{2,t}e^{-\mid x-q_2\mid}+p_1q_{1,t}sgn(x-q_1)e^{-\mid x-q_1\mid}+p_2q_{2,t}sgn(x-q_2)e^{-\mid x-q_2\mid},
\\ m&=2p_1\delta(x-q_1)+2p_2\delta(x-q_2).
\end{split}
\label{tcpd}
\end{eqnarray}
Substituting (\ref{tcp}) and (\ref{tcpd}) into (\ref{cwf2}) with $b=0$, we may obtain the following peakon dynamical system
\begin{eqnarray}
\left\{\begin{array}{l}
p_{1,t}=-\frac{1}{2}k_2p_1p_2 sgn(q_1-q_2)e^{ -\mid q_1-q_2\mid},\\
p_{2,t}=-\frac{1}{2}k_2p_1p_2 sgn(q_2-q_1)e^{ -\mid q_2-q_1\mid},\\
q_{1,t}=-k_1p_1p_2e^{-\mid q_1-q_2\mid}-\frac{1}{3}k_1p_1^2-\frac{1}{2}k_2\left(p_1+p_2e^{ -\mid q_1-q_2\mid}\right),\\
q_{2,t}=-k_1p_1p_2e^{-\mid q_1-q_2\mid}-\frac{1}{3}k_1p_2^2-\frac{1}{2}k_2\left(p_2+p_1e^{ -\mid q_1-q_2\mid}\right).
\end{array}\right. \label{dtcp1}
\end{eqnarray}

For $k_1=0$, $k_2=-2$, (\ref{bCHFFORQ}) recovers the CH equation and (\ref{dtcp1}) is reduced to the two-peakon dynamics of CH equation with $b=0$, which Camassa and Holm studied in \cite{CH}. In particular, the peakon-antipeakon was taken as below
$$u=p_1(t)\left(e^{-\mid x-q_1(t)\mid}-e^{-\mid x+q_1(t)\mid}\right),$$
where
\begin{eqnarray}
p_{1}(t)=c\coth(ct)=-p_2(t),\quad
q_{1}(t)=-\ln\frac{2}{e^{ct}+e^{-ct}}=-q_2(t).
\label{chpq}
\end{eqnarray}
Camassa and Holm \cite{CH} also pointed out that the collision of the peakon-antipeakon occurs at the moment $t=0$. At the moment of collision, the amplitudes $p_1$ and $p_2$ become infinite, but the solution $u$ converges to zero \cite{CH,BC}.

For $k_2=0$ and $b=0$, (\ref{bCHFFORQ}) is exactly the cubic nonlinear equation \cite{OR,Q1,Fo,Fu}
\begin{eqnarray}
m_t=\frac{1}{2}k_1\left[m(u^2-u^2_x)\right]_x, \quad m=u-u_{xx},\label{kqe}
\end{eqnarray}
and the peakon dynamical system (\ref{dtcp1}) is simplified to
\begin{eqnarray}
\left\{\begin{array}{l}
p_{1,t}=0, \quad p_{2,t}=0,\\
q_{1,t}=-k_1p_1p_2e^{-\mid q_1-q_2\mid}-\frac{1}{3}k_1p_1^2,\\
q_{2,t}=-k_1p_1p_2e^{-\mid q_1-q_2\mid}-\frac{1}{3}k_1p_2^2.
\end{array}\right. \label{qetp}
\end{eqnarray}
From the first two equations of (\ref{qetp}), we obtain $p_1(t)=c_1$, $p_2(t)=c_2$, where $c_1,~ c_2$ are two arbitrary constants. For the case of $c_1^2=c_2^2$,
the solutions of $q_1(t)$ and $q_2(t)$ in (\ref{qetp}) are given by
\begin{eqnarray}
\left\{\begin{array}{l}
q_1(t)=-k_1(c_1c_2e^{-|A_1-A_2|}+\frac{1}{3}c_1^2)t+A_1,
\\q_2(t)=-k_1(c_1c_2e^{-|A_1-A_2|}+\frac{1}{3}c_1^2)t+A_2,
\end{array}\right. \label{caseI}
\end{eqnarray}
where $A_1$, $A_2$ are arbitrary integral constants. If $A_1=A_2$, the two-peakon solution is reduced to zero (when $c_2=-c_1$) or a single-peakon (when $c_2=c_1$). If $A_1\neq A_2$, the two-peakon can not collide since $q_1(t)\neq q_2(t)$ for every $t\in (-\infty,+\infty)$. Especially, in the peakon-antipeakon case ($c_1=-c_2$), the collision can't occur for $A_1\neq A_2$. This is very different from the case of the CH equation \cite{CH}. For the case of $c_1^2\neq c_2^2$, we obtain the following solution:
\begin{eqnarray}
\left\{\begin{array}{l}
p_{1}(t)=c_1, \quad p_{2}(t)=c_2,\\
q_{1}(t)=sgn(t)\frac{3k_1c_1c_2}{\mid k_1(c_1^2-c_2^2)\mid}\left(e^{-\mid \frac{1}{3}k_1(c_1^2-c_2^2)t\mid}-1\right)-\frac{1}{3}k_1c_1^2t,\\
q_{2}(t)=sgn(t)\frac{3k_1c_1c_2}{\mid k_1(c_1^2-c_2^2)\mid}\left(e^{-\mid \frac{1}{3}k_1(c_1^2-c_2^2)t\mid}-1\right)-\frac{1}{3}k_1c_2^2t.
\end{array}\right. \label{qepq}
\end{eqnarray}
The two-peakon collision occurs at the moment $t=0$, since $q_1(0)=q_2(0)=0$. Without loss of generality, let us suppose $c_1^2<c_2^2$. From the formula (\ref{qepq}), we know that for $t<0$, the tall and fast peakon (with the amplitude $|c_2|$ and peak position $q_2$) chases after the short and slow peakon (with the  amplitude $|c_1|$ and peak position $q_1$). At the moment $t=0$, the two-peakon collides and overlaps into the peakon: $u(x,0)=(c_1+c_2)e^{-|x|}$. After the collision ($t>0$), the two-peakon departs, and the tall and fast peakon surpasses the short and slow one.
See Figures 1 and 2 for the developments of this kind of two-peakon.

{\bf Remark 1.} Our results show that the collision of two-peakon of equation (\ref{kqe}) is very different from the case of CH equation.
In the peakon-antipeakon case, the collision can occur for the CH equation, but could not happen for equation (\ref{kqe}).
The collision of two-peakon of equation (\ref{kqe}) occurs in the ``chase'' case as shown above (see Figures 1 and 2).

$ $\\

Next, we want to derive the solutions of (\ref{dtcp1}) for the general case $k_1\neq 0$ and $k_2\neq 0$. To do so, let us introduce the transformations $p(t)=p_1(t)+p_2(t)$, $q(t)=q_1(t)+q_2(t)$, $P(t)=p_1(t)-p_2(t)$, $Q(t)=q_1(t)-q_2(t)$.
Then (\ref{dtcp1}) is transformed to
\begin{eqnarray}
\left\{\begin{array}{l}
p_{t}=0,\\
q_{t}=-\frac{1}{2}k_1(p^2-P^2)e^{ -\mid Q\mid}-\frac{1}{6}k_1(p^2+P^2)-\frac{1}{2}k_2p(1+e^{ -\mid Q\mid}),\\
P_{t}=-\frac{1}{4}k_2(p^2-P^2)sgn(Q) e^{-\mid Q\mid},\\
Q_{t}=\frac{1}{2}k_2P(e^{-\mid Q\mid}-1)-\frac{1}{3}k_1pP.
\end{array}\right. \label{dtcp2}
\end{eqnarray}
From the first equation of (\ref{dtcp2}), we know that $p(t)=A_1$, where $A_1$ is an arbitrary constant.
Let $\Gamma=1+\frac{2}{3}\frac{k_1}{k_2}A_1$,
then we have the following results.

{\bf Case 1.} If $0<\Gamma\leq 1$, then (\ref{dtcp2}) admits the following solution
\begin{eqnarray}
\begin{split}
P(t)&=\pm a_2\frac{1+\tilde A_3e^{Bt}}{1-\tilde A_3e^{Bt}}, \quad
Q(t)=\pm\ln\frac{4\Gamma a_2^2\tilde A_3e^{Bt}}{a_2^2(1+\tilde A_3e^{Bt})^2-A_1^2(1-\tilde A_3e^{Bt})^2},
\\
q(t)&=-\ln\frac{\mid \tilde A_3e^{Bt}-\frac{A_1+a_2}{A_1-a_2}\mid}{\mid \tilde A_3e^{Bt}-\frac{A_1-a_2}{A_1+a_2}\mid}-\frac{2k_1a_2^2(3\Gamma-1)}{3B(\tilde A_3e^{Bt}-1)}
-\frac{1}{2}[k_2A_1+\frac{1}{3}k_1(A_1^2+a_2^2)]t+D,
\end{split}
\label{tpc21}
\end{eqnarray}
where $a_2>|A_1|$, $\tilde A_3>0$, $B=-\frac{1}{2}a_2k_2\Gamma$, and $D$ is an arbitrary constant.

{\bf Case 2.} If $\Gamma>1$, then (\ref{dtcp2}) has the following solution
\begin{eqnarray}
\begin{split}
P(t)&=\pm a_2\frac{1-\tilde A_3e^{Bt}}{1+\tilde A_3e^{Bt}}, \quad Q(t)=\pm\ln\frac{-4\Gamma a_2^2\tilde A_3e^{Bt}}{a_2^2(1-\tilde A_3e^{Bt})^2-A_1^2(1+\tilde A_3e^{Bt})^2},
\\
q(t)&=-\ln\frac{\mid \tilde A_3e^{Bt}+\frac{A_1+a_2}{A_1-a_2}\mid}{\mid \tilde A_3e^{Bt}+\frac{A_1-a_2}{A_1+a_2}\mid}+\frac{2k_1a_2^2(3\Gamma-1)}{3B(\tilde A_3e^{Bt}+1)}
-\frac{1}{2}[k_2A_1+\frac{1}{3}k_1(A_1^2+a_2^2)]t+D,
\end{split}
\label{tpc1}
\end{eqnarray}
where $0<a_2<\frac{|A_1|}{\sqrt{\Gamma}}$, $\tilde A_3>0$, $B=-\frac{1}{2}a_2k_2\Gamma$, and $D$ is an arbitrary constant.

Let us now give special two-peakon solutions based on Case 1 and Case 2.

{\bf For Case 1,} let $A_1=0$, $a_2=2$, $\tilde A_3=1$, $k_1=k_2=-2$, and $D=0$, then we have
\begin{eqnarray}
p(t)=0, \quad P(t)=2\coth(t), \quad q(t)=\frac{16}{3(e^{2t}-1)}+\frac{4}{3}t, \quad Q(t)=-\ln\frac{4e^{2t}}{(1+e^{2t})^2}.
\label{pq1}
\end{eqnarray}
These yield
\begin{eqnarray}
\begin{split}
p_1(t)&=\coth(t), \quad q_1(t)=\frac{8}{3(e^{2t}-1)}+\ln(e^{2t}+1)-\frac{1}{3}t-\ln2,
\\ p_2(t)&=-\coth(t), \quad q_2(t)=\frac{8}{3(e^{2t}-1)}-\ln(e^{2t}+1)+\frac{5}{3}t+\ln2.
\end{split}
\label{spq}
\end{eqnarray}
Therefore, we obtain the following 
peakon-antipeakon solution 
\begin{eqnarray}
u(x,t)=\coth(t)\left(e^{-\mid x-q_1(t)\mid}-e^{-\mid x-q_2(t)\mid}\right),\label{tcpu2}
\end{eqnarray}
where $q_1(t)$ and $q_2(t)$ are shown in (\ref{spq}).
From (\ref{pq1}), one can easily know that the collision occurs at the moment $t=0$ since $Q(0)=0$. From (\ref{spq}), we may compute
\begin{eqnarray}
\lim_{t\rightarrow 0} p_1(t)=-\lim_{t\rightarrow 0} p_2(t)=\infty, \quad \lim_{t\rightarrow 0} q_1(t)=\lim_{t\rightarrow 0} q_2(t)=\infty.
\label{pqd}
\end{eqnarray}
But from (\ref{tcpu2}), we may infer that
\begin{eqnarray}
\lim_{t\rightarrow 0}u(x,t)=0, ~~ {\rm for ~~ every} ~~ x\in R,
\label{uv}
\end{eqnarray}
which indicates that the peakon and the antipeakon vanish when they overlap.
Guided by the above results, we may describe the dynamics of peakon-antipeakon solution (\ref{tcpu2}) as follows. For $t<0$, the peak is at $q_2(t)$ and the trough is at $q_1(t)$. The peak and the trough approach each other as $t$ goes to $0$. At the moment $t=0$, the peakon and the antipeakon collide and vanish. After their collision ($t>0$), they depart and redevelop with the trough at $q_2(t)$ and the peak at $q_1(t)$. Figure 3 shows the peakon-antipeakon dynamics.

{\bf Remark 2.} The amplitudes $p_1(t)$, $p_2(t)$ in formula (\ref{spq}) are the same as those of the CH equation \cite{CH}, but the peak positions $q_1(t)$, $q_2(t)$ are different (see formulas (\ref{chpq}) and (\ref{spq})). In the CH equation, only $p_1(t)$ and $p_2(t)$ become infinite at the
instant of collision \cite{CH,CH2}. In the new equation (\ref{bCHFFORQ}) with $b=0$, not only $p_1(t)$, $p_2(t)$ but also $q_1(t)$ and $q_2(t)$ become infinite at the
instant of collision (see formula (\ref{pqd})).
However, in both case, the peakon-antipeakon vanishes when the overlap occurs (see formula (\ref{uv})).

$ $\\

{\bf For Case 2,} choosing $k_1=k_2=-2$, $A_1=3$, $\tilde A_3=1$, $\Gamma=3$, $a_2=\sqrt{3}$, $B=3\sqrt{3}$ forces the first two equations in (\ref{tpc1}) to
\begin{eqnarray}
P(t)=\sqrt{3}sgn(t)\frac{1-e^{3\sqrt{3}t}}{1+e^{3\sqrt{3}t}},
\quad
Q(t)=sgn(t)\ln\frac{6e^{3\sqrt{3}t}}{e^{6\sqrt{3}t}+4e^{3\sqrt{3}t}+1}.
\label{tpc12}
\end{eqnarray}
Substituting (\ref{tpc12}) into the second equation of (\ref{dtcp2}), we may obtain the expression of $q(t)$ and finally arrive at the solution
\begin{eqnarray}
\begin{split}
p_1(t)&=\frac{3}{2}-\frac{\sqrt{3}}{2}sgn(t)\tanh(\frac{3\sqrt{3}t}{2}), \quad
p_2(t)=\frac{3}{2}+\frac{\sqrt{3}}{2}sgn(t)\tanh(\frac{3\sqrt{3}t}{2}),
\\ q_1(t)&=\frac{1}{2}sgn(t)\ln\frac{6e^{3\sqrt{3}t}}{e^{6\sqrt{3}t}+4e^{3\sqrt{3}t}+1}
-\frac{1}{2}\ln\frac{e^{3\sqrt{3}t}+2+\sqrt{3}}{e^{3\sqrt{3}t}+2-\sqrt{3}}
-\frac{16\sqrt{3}}{9(e^{3\sqrt{3}t}+1)}+\frac{7}{2}t,
\\ q_2(t)&=-\frac{1}{2}sgn(t)\ln\frac{6e^{3\sqrt{3}t}}{e^{6\sqrt{3}t}+4e^{3\sqrt{3}t}+1}
-\frac{1}{2}\ln\frac{e^{3\sqrt{3}t}+2+\sqrt{3}}{e^{3\sqrt{3}t}+2-\sqrt{3}}
-\frac{16\sqrt{3}}{9(e^{3\sqrt{3}t}+1)}+\frac{7}{2}t.
\end{split}
\label{spq2}
\end{eqnarray}
The formulas (\ref{tpc12}) show that the collision happens at the moment $t=0$ since $Q(0)=0$. From (\ref{spq2}), we know that the two-peakon collides and overlaps into the peakon $u(x,0)=3e^{-|x+\frac{1}{2}\ln(2+\sqrt{3})+\frac{8\sqrt{3}}{9}|}$ at the moment $t=0$. After the collision, the two-peakon departs and redevelops. See Figure 4 for the profile of the two-peakon dynamics.

\subsection{N-peakon dynamical systems}
In general, we suppose that $N$-peakon has the following form
\begin{eqnarray}
u(x,t)=\sum_{j=1}^N p_j(t)e^{-\mid x-q_j(t)\mid}.\label{Ncp}
\end{eqnarray}
Similar to the cases of one-peakon and two-peakon, but with a lengthy calculation, we are able to obtain the following $N$-peakon dynamical system
\begin{eqnarray}
\begin{split}
p_{j,t}&=-\frac{1}{2}k_2p_j\sum_{k=1}^Np_k sgn(q_j-q_k)e^{ -\mid q_j-q_k\mid},\\
q_{j,t}&=-\frac{1}{2}k_2\sum_{k=1}^Np_k e^{ -\mid q_j-q_k\mid}+\frac{1}{2}k_1\left(\frac{1}{3}p_j^2-\sum_{i,k=1}^N  p_ip_k(1-sgn(q_j-q_i)sgn(q_j-q_k))e^{ -\mid q_j-q_i\mid-\mid q_j-q_k\mid}\right).
\end{split}   \label{dNcp}
\end{eqnarray}
Different from the $N$-peakon system of the CH equation, the above system can not directly be rewritten in the standard form of a canonical Hamiltonian system. It is very interesting to see whether the above system is an integrable Hamiltonian
system under a Poisson structure. We will investigate this in the near future.

\begin{figure}
\begin{minipage}[t]{0.5\linewidth}
\centering
\includegraphics[width=2.2in]{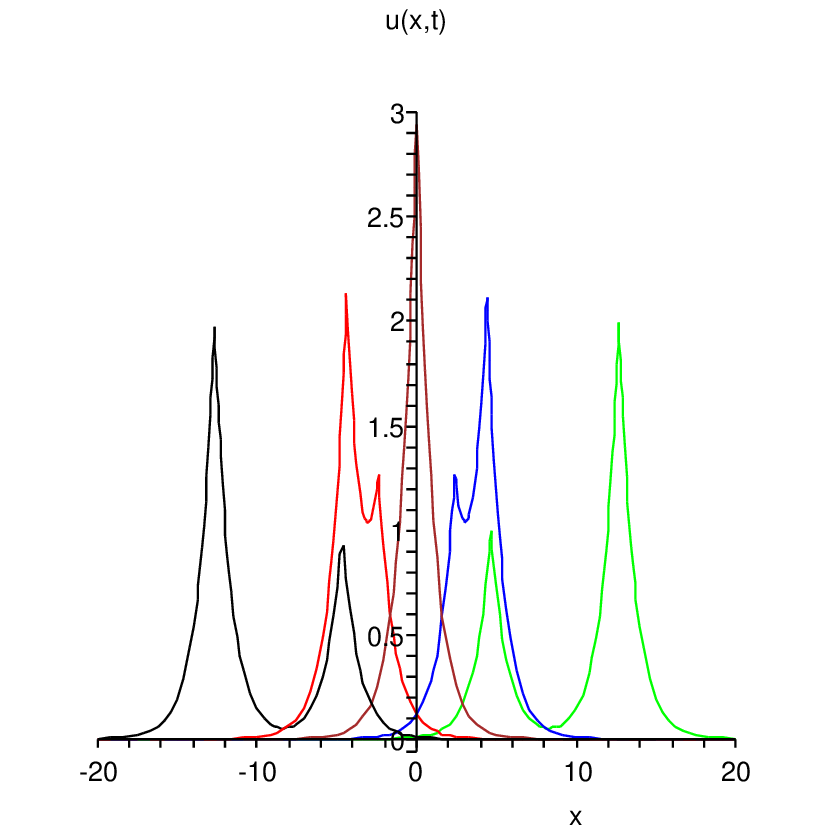}
\caption{\small{The two-peakon solution determined by (\ref{qepq}) with $c_1=1$, $c_2=2$, $k_1=-2$. Black line: $t=-4$; Red line: $t=-1$; Brown line: $t=0$ (collision); Blue line: $t=1$; Green line: $t=4$.}}
\label{fig:side:a}
\end{minipage}%
\hspace{2.0ex}
\begin{minipage}[t]{0.5\linewidth}
\centering
\includegraphics[width=2.2in]{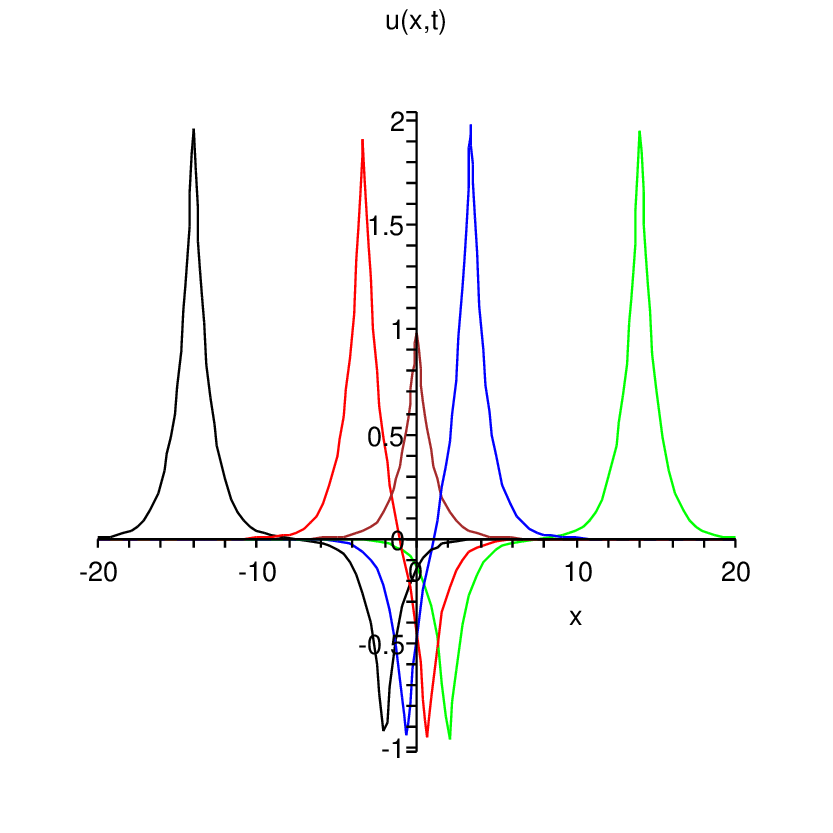}
\caption{\small{ The two-peakon solution determined by (\ref{qepq}) with $C=-1$, $c_2=2$, $k_1=-2$. Black line: $t=-6$; Red line: $t=-2$; Brown line: $t=0$ (collision); Blue line: $t=2$; Green line: $t=6$.}}
\label{fig:side:b}
\end{minipage}
\begin{minipage}[t]{0.5\linewidth}
\centering
\includegraphics[width=2.2in]{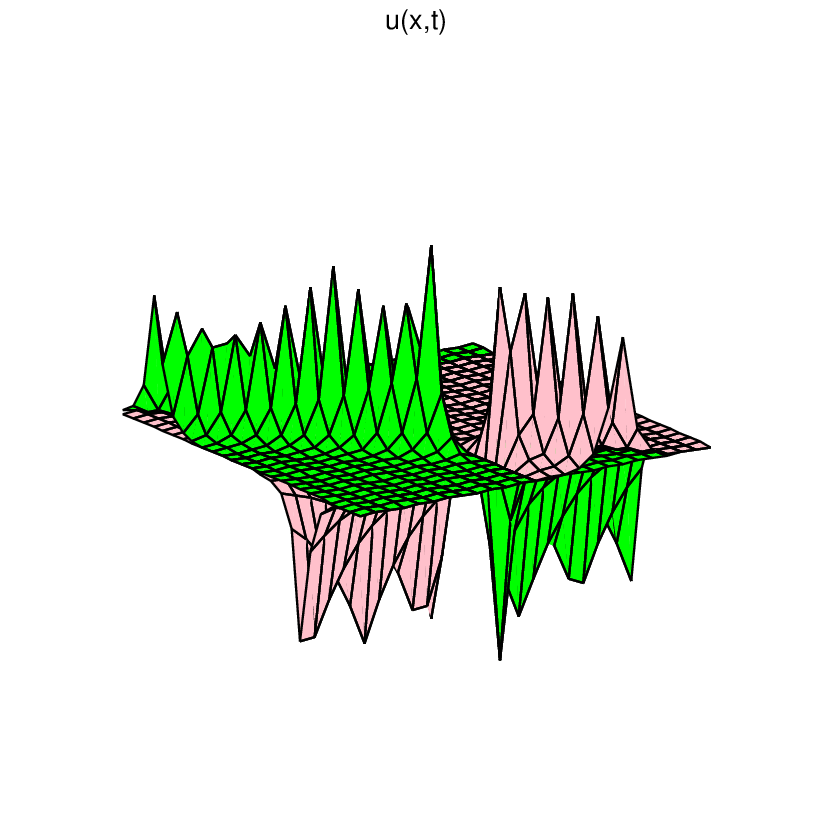}
\caption{\small{The peakon-antipeakon solution (\ref{tcpu2}). Pink: peakon (and antipeakon) with the peak
(and trough) position $q_2$; Green: antipeakon (and peakon) with the trough (and peak) position $q_1$.}}
\label{fig:side:b}
\end{minipage}
\hspace{2.0ex}
\begin{minipage}[t]{0.5\linewidth}
\centering
\includegraphics[width=2.2in]{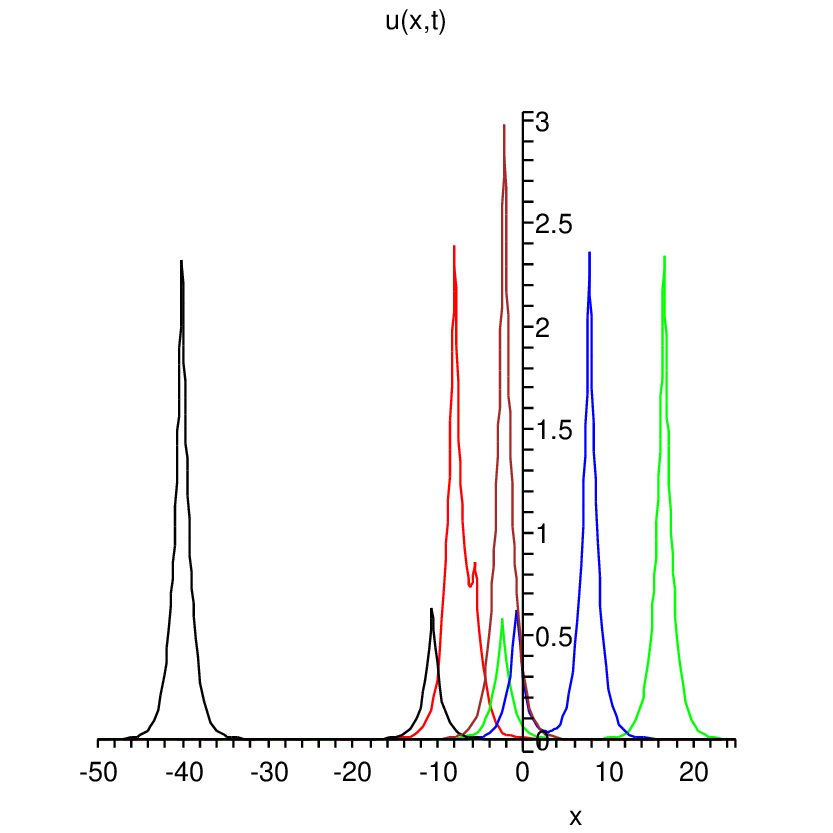}
\caption{\small{The solution $u(x,t)$ with $p_1$, $p_2$ and
$q_1$, $q_2$ shown in (\ref{spq2}). Black line: $t=-6$; Red line: $t=-0.8$; Brown line: $t=0$ (collision); Blue line: $t=2$; Green line: $t=4$.}}
\label{fig:side:b}
\end{minipage}
\begin{minipage}[t]{0.5\linewidth}
\centering
\includegraphics[width=2.2in]{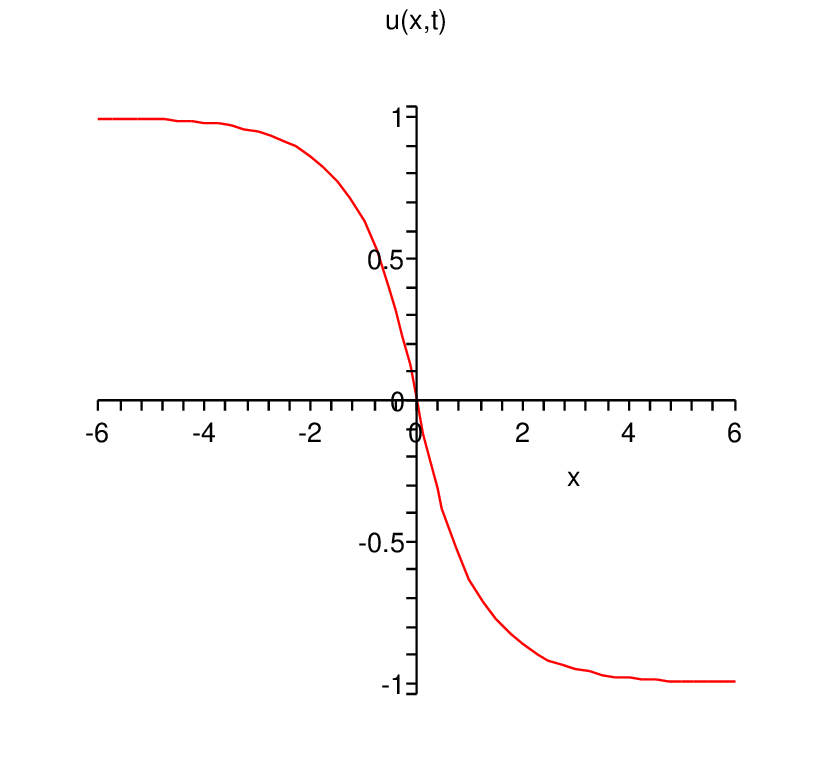}
\caption{\small{The kink solution (\ref{kink1s}) for $t=0$.}}
\label{fig:side:a}
\end{minipage}
\hspace{2.0ex}
\begin{minipage}[t]{0.5\linewidth}
\centering
\includegraphics[width=2.2in]{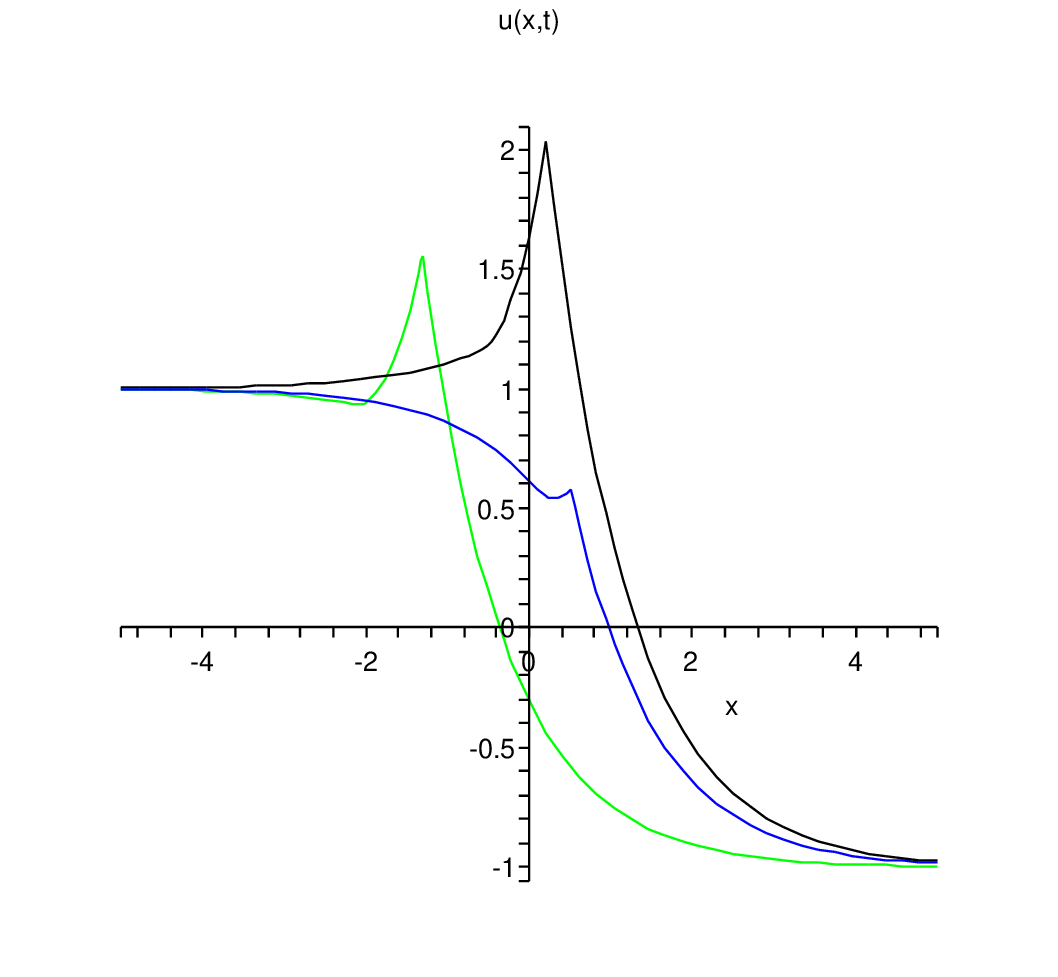}
\caption{\small{The kink-peakon interacted solution. Black line: $t=2$; Blue line: $t=1$; Green line: $t=0$.}}
\label{fig:side:b}
\end{minipage}
\end{figure}

\section{Weak kink solution in the case $b\neq0$}

We have already shown that equation (\ref{bCHFFORQ}) admits peakon solutions in the case $b=0$. It is natural to ask what kind of solution one may obtain for the case $b\neq 0$. Here we will reveal that equation (\ref{bCHFFORQ}) with $k_2=0$ and $b\neq0$ (i.e. cubic nonlinear equation (\ref{qe}) with $b\neq 0$) possesses the weak kink and kink-peakon interacted solutions.

\subsection{Single weak kink solution}
We seek the solution of equation (\ref{bCHFFORQ}) in the form of
\begin{eqnarray}
u=Csgn(x-ct)\left(e^{-\mid x-ct\mid}-1\right),\label{kink1}
\end{eqnarray}
where the constant $C$ is to be determined. If $C\neq0$, (\ref{kink1}) actually is a kink solution since
\begin{eqnarray}
\lim_{x\rightarrow +\infty} u=-\lim_{x\rightarrow -\infty} u=-C.
\label{reasonkink}
\end{eqnarray}
One may easily check that the first order partial derivatives of (\ref{kink1}) read
\begin{eqnarray}
u_x=-Ce^{-\mid x-ct\mid}, \quad u_t=cCe^{-\mid x-ct\mid}.\label{dkink1}
\end{eqnarray}
The second order partial derivatives of (\ref{kink1}) do not exist at $x=ct$.
Therefore, like the case of peakon solutions, the kink solution in the form of (\ref{kink1}) should also be understood in the weak sense.

Substituting (\ref{kink1}) and (\ref{dkink1}) into the weak form (\ref{cwf2}), we arrive at
\begin{eqnarray}
cCe^{-\mid x-ct\mid}=\frac{1}{2}[k_2C^2(x-ct)-bC]e^{-\mid x-ct\mid}-\frac{1}{2}[k_1C^3+bC](x-ct)sgn(x-ct)e^{-\mid x-ct\mid}.\label{1ks}
\end{eqnarray}
If two sides of equation (\ref{1ks}) match, we should require
\begin{eqnarray}
\left\{\begin{array}{l}
k_2=0,\\
-\frac{1}{2}bC-cC=0,\\
k_1C^3+bC=0,
\end{array}\right. \label{ikss}
\end{eqnarray}
which leads to
\begin{eqnarray}
\left\{\begin{array}{l}
k_2=0,\\
c=-\frac{1}{2}b,\\
C=\pm \sqrt{\frac{-b}{k_1}}.
\end{array}\right. \label{ikss}
\end{eqnarray}

{\bf Remark 3.}  Formula (\ref{ikss}) shows that equation (\ref{bCHFFORQ}) possesses the weak kink solution in the form (\ref{kink1}) for the case of $k_2=0$ and $b\neq0$. In formula (\ref{ikss}), $k_2=0$ implies that the CH equation (\ref{bCH}) can not allow a weak kink solution in the form of (\ref{kink1}). $c=-\frac{1}{2}b$ means the kink wave speed is exactly $-\frac{1}{2}b$. This is very different from the single-peakon solution (the wave speed in the single-peakon (\ref{ocp}) is an arbitrary constant $c$).

 $ $\\

In particular, we take $k_2=0$, $k_1=-b=2$, then equation (\ref{bCHFFORQ}) is cast to
\begin{eqnarray}
m_t+2u_x-\left[m(u^2-u^2_x)\right]_x=0, \quad m=u-u_{xx},\label{2qe}
\end{eqnarray}
and the corresponding weak kink solution is
\begin{eqnarray}
u=sgn(x-t)\left(e^{-\mid x-t\mid}-1\right).\label{kink1s}
\end{eqnarray}
See Figure 5 for the profile of this kink wave solution.

\subsection{Kink-peakon interacted solutions}
Let us first point out that equation (\ref{bCHFFORQ}) with $b\neq0$ does not allow the two-kink solution in the form of the superposition of two single-kink solutions:
\begin{eqnarray}
u=p_1(t)sgn(x-q_1(t))\left(e^{-\mid x-q_1(t)\mid}-1\right)+p_2(t)sgn(x-q_2(t))\left(e^{-\mid x-q_2(t)\mid}-1\right).
\label{tk}
\end{eqnarray}
In fact, substituting (\ref{tk}) into the weak form (\ref{cwf2}) of equation (\ref{bCHFFORQ}), we find that the solution assumed in the form (\ref{tk}) is reduced to zero or single-kink solution (\ref{kink1}).

Instead of considering the solution in the form (\ref{tk}), let us make the following ansatz of solution
\begin{eqnarray}
u=p_1(t)sgn(x-q_1(t))\left(e^{-\mid x-q_1(t)\mid}-1\right)+p_2(t)e^{-\mid x-q_2(t)\mid},
\label{tcpk}
\end{eqnarray}
which actually describes {\bf a new phenomena of kink-peakon interacted dynamics} in soliton theory.
Apparently, with the help of distribution, $u_x$ and $u_t$ can be calculated below:
\begin{eqnarray}
\begin{split}
u_x&=-p_1e^{-\mid x-q_1\mid}-p_2sgn(x-q_2)e^{-\mid x-q_2\mid},\\
u_t&=p_{1,t}sgn(x-q_1(t))\left(e^{-\mid x-q_1(t)\mid}-1\right)+p_{2,t}e^{-\mid x-q_2\mid}+p_1q_{1,t}e^{-\mid x-q_1\mid}+p_2q_{2,t}sgn(x-q_2)e^{-\mid x-q_2\mid}.
\end{split}
\label{tcpdk}
\end{eqnarray}
Substituting (\ref{tcpk}) and (\ref{tcpdk}) into (\ref{cwf2}) with $k_2=0$, we arrive at
\begin{eqnarray}
\left\{\begin{array}{l}
p_{1}=\pm\sqrt{\frac{-b}{k_1}},\\
p_{2,t}=k_1p_1^2p_2 sgn(q_2-q_1)e^{ -\mid q_1-q_2\mid},\\
q_{1,t}=-\frac{1}{2}b-k_1p_1p_2sgn(q_2-q_1)e^{-\mid q_1-q_2\mid},\\
q_{2,t}=-\frac{1}{3}k_1p_2^2-\frac{1}{2}k_1p_1^2+k_1\left(p_1^2-p_1p_2sgn(q_2-q_1)\right)e^{-\mid q_1-q_2\mid}+k_1sgn(q_2-q_1)p_1p_2.
\end{array}\right. \label{kink2}
\end{eqnarray}
Let us choose $k_1=-b=2$, then $p_1=\pm 1$. Without loss of generality, taking $p_1=1$ forces (\ref{kink2}) to
\begin{eqnarray}
\left\{\begin{array}{l}
p_{2,t}=2p_2sgn(q_2-q_1) e^{ -\mid q_1-q_2\mid},\\
q_{1,t}=1-2p_2sgn(q_2-q_1)e^{-\mid q_1-q_2\mid},\\
q_{2,t}=-\frac{2}{3}p_2^2-1+2\left(1-p_2sgn(q_2-q_1)\right)e^{-\mid q_1-q_2\mid}+2p_2sgn(q_2-q_1).
\end{array}\right. \label{kink2r}
\end{eqnarray}
To solve the above system, let us make an assumption $q_1<q_2$. Therefore, integrating equation (\ref{kink2r}), we may obtain
\begin{eqnarray}
\left\{\begin{array}{l}
q_1=t-p_2+A_1,
\\
q_2=t-p_2-\ln(\frac{1}{9}p_2^2-\frac{1}{2}p_2+1+\frac{A_2}{2p_2})+A_1,
\\
p_{2,t}=\frac{2}{9}p_2^3-p_2^2+2p_2+A_2,
\end{array}
\right.
\label{k2r1}
\end{eqnarray}
where $A_1$ and $A_2$ are two arbitrary constants.
Letting $A_2=0$, then we may solve the third equation of (\ref{k2r1}) for $p_2$, which has the following  implicit form:
\begin{eqnarray}
\ln \left|p_2\right|-\frac{1}{2}\ln(p_2^2-\frac{9}{2}p_2+9)+\frac{3\sqrt{7}}{7}\arctan(\frac{4p_2-9}{3\sqrt{7}})=2t+A_3.
\label{k2r2}
\end{eqnarray}
See Figure 6 for the profile of the kink-peakon interactional solution with $A_1=A_2=A_3=0$. Here, we do not have more further explicit results about
the collisions between kinks and peakons due to the implicit form of the kink-peakon solution.

In general, we may make the following ansatz of the solution for equation (\ref{bCHFFORQ}) with $k_2=0$ and $b\neq0$:
\begin{eqnarray}
u=p_0(t)sgn(x-q_0(t))\left(e^{-\mid x-q_0(t)\mid}-1\right)+\sum_{j=1}^N p_j(t)e^{-\mid x-q_j(t)\mid},\label{nss}
\end{eqnarray}
which can be viewed as the interaction of single-kink and $N$-peakon solution.
Through a very lengthy calculation, we are able to obtain the following interactional dynamical system of single-kink and $N$-peakon:
\begin{eqnarray}
\left\{\begin{split}
p_0&=\pm \sqrt{-\frac{b}{k_1}},\\
q_{0,t}&=\frac{1}{2}k_1p_0^2+k_1p_0\sum_{i=1}^Np_i sgn(q_0-q_i)e^{ -\mid q_0-q_i\mid}
\\&+\frac{1}{2}k_1\sum_{i,k=1}^Np_ip_k sgn(q_i-q_k)(sgn(q_k-q_0)-sgn(q_i-q_0))e^{ -\mid q_i-q_k\mid},\\
p_{j,t}&=k_1p_0^2p_j sgn(q_j-q_0)e^{-\mid q_0-q_j\mid}+k_1p_0p_j\sum_{i=1}^Np_isgn(q_j-q_i)sgn(q_j-q_0) e^{ -\mid q_j-q_i\mid},\\
q_{j,t}&=\frac{1}{6}k_1p_j^2-\frac{1}{2}k_1p_0^2(1-2e^{-\mid q_0-q_j\mid})
-\frac{1}{2}k_1\sum_{i,k=1}^N  p_ip_k(1-sgn(q_j-q_i)sgn(q_j-q_k))e^{ -\mid q_j-q_i\mid-\mid q_j-q_k\mid}
\\&-k_1p_0\sum_{i=1}^N  p_i\left(sgn(q_j-q_0)(e^{-\mid q_0-q_j\mid}-1)e^{-\mid q_i-q_j\mid}-sgn(q_j-q_i)e^{-\mid q_0-q_j\mid-\mid q_i-q_j\mid}\right).
\end{split}\right.
\label{Ncp}
\end{eqnarray}
The above system is not presented in the canonical Hamiltonian system. We still do not know  whether this system is integrable 
under a Poisson structure.


\section{Smooth soliton solutions}

In this section, we study all possible single smooth soliton solutions to
equation (\ref{bCHFFORQ}). To do so, let $u(x,t)=\phi(x-ct)=\phi(\xi)$, where $c$ is the wave speed.
Substituting  it into equation (\ref{bCHFFORQ}) and integrating it, 
we have
\begin{eqnarray} \phi^{\prime\prime}\left(c+\frac12k_2\phi+\frac12k_1\phi^2-\frac12k_1(\phi^{\prime})^2\right)=\left((b+c)\phi+\frac34k_2\phi^2+\frac12k_1\phi^3\right)
-\left(\frac14k_2+\frac12k_1\phi\right)(\phi^{\prime})^2,\label{Int}
\end{eqnarray}
where $"\prime" $ stands for the derivative with respect to $\xi$.
Equation (\ref{Int}) can be rewritten as the following two-dimensional system
\begin{eqnarray}
\frac{d\phi}{d\xi}=y,\ \ \frac{dy}{d\xi}=\frac{-\left(\frac14k_2+\frac12k_1\phi\right)
y^2+\phi((b+c)+\frac34k_2\phi+\frac12k_1\phi^2)}{c+\frac12k_2\phi+\frac12k_1\phi^2-\frac12k_1y^2
}, \label{Int2D}
\end{eqnarray}
which has the following first integral
\begin{eqnarray}
H(\phi,y)=\frac12
y^2\left(c+\frac12k_2\phi+\frac12k_1\phi^2\right)-
\frac18k_1y^4-\left((b+c)\phi^2+\frac14k_2\phi^3+\frac18k_1\phi^4\right)=h. \label{IntHamil}
\end{eqnarray}

Without loss of generality, we assume that the wave speed $c$ is a positive constant.
Then, system (\ref{Int2D}) is a three-parameter planar dynamical system
with the triple tuple $(b, k_1, k_2)$. We only pay attention to the
bounded solutions of (\ref{Int2D}).

We notice that for $k_1\neq 0$, the right hand of the second
equation in (\ref{Int2D}) is not continuous on the hyperbola
$c+\frac12k_2\phi+\frac12k_1\phi^2-\frac12k_1y^2=0$, i.e.,
$\left(\phi+\frac{k_2}{2k_1}\right)^2-y^2=\frac{k_2^2-8k_1c}{4k_1^2}.$
In other words,
$\phi_{\xi}^{\prime\prime}$ is not well-defined on this curve in the phase plane $(\phi,y)$.
System  (\ref{Int2D}) is
called the second kind of singular travelling wave system \cite{LJB1}.

Let us assume   $c>0, k_1\neq 0.$ Imposing the transformation
$d\xi=(c+\frac12k_2\phi+\frac12k_1\phi^2-\frac12k_1y^2)d\zeta$ for
$c+\frac12k_2\phi+\frac12k_1\phi^2-\frac12k_1y^2\neq 0$ on system
 (\ref{Int2D}) leads to the following cubic system
\begin{eqnarray}
\left\{\begin{array}{ll}
&\frac{d\phi}{d\zeta}=y(c+\frac12k_2\phi+\frac12k_1\phi^2-\frac12k_1y^2),\\[1ex]
&\frac{dy}{d\zeta}=-\left(\frac14k_2+\frac12k_1\phi\right)
y^2+\phi((b+c)+\frac34k_2\phi+\frac12k_1\phi^2),\end{array}\right.
\label{cubic}
\end{eqnarray}
which can be rewritten as
\begin{eqnarray}
\left\{\begin{array}{ll}
&\frac{d\varphi}{d\eta}=y(\varphi^2-y^2-R),\\[1ex]
&\frac{dy}{d\eta}=-\varphi y^2+\varphi^3+A\varphi+B=-\varphi
y^2+\left(\varphi-\frac{k_2}{2k_1}\right)\left(\varphi^2+\frac{k_2}{2k_1}\varphi+\frac{4k_1(b+c)-k_2^2}{2k_1^2}\right),\end{array}\right.
\label{newcubic}\end{eqnarray}
where $\phi=\varphi-\frac{k_2}{2k_1},
\eta=\frac{k_1}{2}\zeta,$ $R=\frac{k_2^2-8k_1c}{4k_1^2}$, $A=\frac{8k_1(b+c)-3k_2^2}{4k_1^2},$ and $
B=\frac{k_2(k_2^2-4k_1(b+c))}{4k_1^3}.$ Apparently, (\ref{newcubic}) has the following first integral:
\begin{eqnarray}
H_1(\varphi,y)=-\frac14(\varphi^2-y^2)^2-\frac12A\varphi^2-\frac12Ry^2-B\varphi=h. \label{H1}
\end{eqnarray}
Thus, we have
\begin{eqnarray}
y^2=(\varphi^2-R)\pm\sqrt{(R^2-4h)-4B\varphi-2(A+R)\varphi^2}. \label{yequ}
\end{eqnarray}

To investigate the exact soliton solutions of equation (\ref{bCHFFORQ}),
let us consider system (\ref{newcubic}) with $B=0,$ namely homogeneous system.


{\bf 1. M-shape and W-shape soliton solutions}

Assume $k_1<0, k_2^2=4k_1(b+c)>0, b+c<0,$ and let
$\omega_1=\frac{1}{2|k_1|}\sqrt{b^2-c^2}.$ Then, system (\ref{newcubic}) is able to be solved with
the following smooth M-shape and W-shape soliton solution for equation (\ref{bCHFFORQ}):
\begin{equation}\label{16}
  \begin{array}{ll}
&\varphi(\chi)=\pm\frac{(b-c)\sqrt{\frac{2(b+c)}{k_1}}\cosh(\chi)}{(b+c)-2c\cosh^2(\chi)},\\[1ex]
&\xi(\chi)=-\frac12\sqrt{\frac{b-c}{b+c}}\chi-\ln\left(\frac{1+\sqrt{\frac{b+c}{b-c}}\tanh(\chi)}{1-\sqrt{\frac{b+c}{b-c}}\tanh(\chi)}\right).\end{array}
\end{equation}

When $-\infty<b<-3c$, 
the functions defined by 
(\ref{16}) take on the graph of smooth M-shape and
W-shape profiles 
shown in Fig.7 (7-1).
When $-3c\leq b<-c$, the
functions defined by (\ref{16}) give smooth   solitons
shown in Fig.7 (7-2). 

\begin{center}
\begin{tabular}{cc}
 \epsfxsize=3.5cm\epsffile{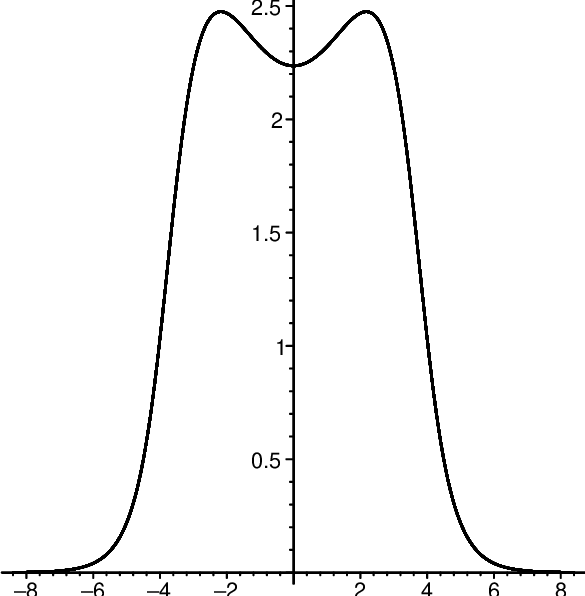} &
\epsfxsize=3.5cm \epsfysize=3.5cm \epsffile{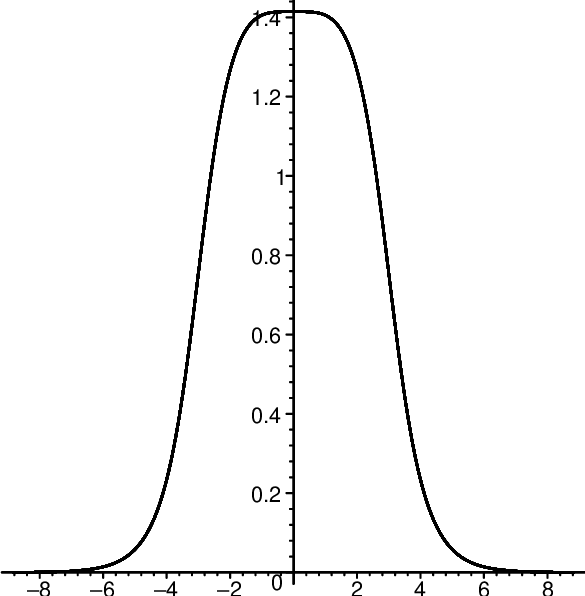}\\
(7-1) $\varphi(\xi).$ & 
(7-2) $\varphi(\xi).$\\
\multicolumn{2}{c}{Fig.7 \   The profiles of  waves with respect to $\xi$} \\
\end{tabular}
\end{center}

{\bf 2. Smooth kink and anti-kink  wave solutions}

Suppose $k_1>0, k_2^2=4k_1(b+c)>0, b+c>0, b<0, c>0,$ and let $\alpha_2=\sqrt{\frac{|b|}{b+c}},\
\omega_2=\frac{2}{k_1}\sqrt{c(b+c)}.$
Then, system (\ref{newcubic}) may be solved with
the following
kink  and anti-kink wave solutions for equation (\ref{bCHFFORQ}):
\begin{equation}\label{19}
  \begin{array}{ll}
&\varphi(\chi)=\pm\frac{\sqrt{|b|(b+c)}\sinh(\chi)}{(\sqrt{c}+\sqrt{|b|}\cosh(\chi)},\\[1ex]
&\xi(\chi)=2c\chi-\frac{1}{4k_1}\ln\frac{1+2w+w^2}{1-2w+w^2}
+\sqrt{c(c+b)}\ln\frac{(w+1)(\alpha_2^2+w+\alpha_2\sqrt{1+\alpha_2^2}\cosh\chi)}{(w-1)(\alpha_2^2-w+\alpha_2\sqrt{1+\alpha_2^2}\cosh\chi)},\end{array}
\end{equation}
where $w=\alpha_2\sinh(\chi).$

However in the case of $k_1>0, k_2^2=4k_1(b+c)>0, b=0,$
we can solve the system (\ref{newcubic}) with
the following kink and anti-kink wave solutions for system (\ref{bCHFFORQ}) in a simpler form:
\begin{equation}\label{21}
  \begin{array}{ll}
&\varphi(\chi)=\frac{\sqrt{\frac{c}{k_1}}}{1+e^{-\chi}},\\[1ex]
&\xi(\chi)=\pm\frac12(\chi+\ln\cosh\chi).\end{array}
\end{equation}

{\bf 3. Smooth soliton solutions}

Suppose $k_1>0, k_2^2=4k_1(b+c)>0, 0<b<c,$ and let
 $\alpha_3=\sqrt{\frac{b}{c-b}}$, and $\omega_3=\frac{2}{k_1}\sqrt{c(c-b)}.$
Then, we can solve the system (\ref{newcubic}) with
the following soliton solutions for equation (\ref{bCHFFORQ})
\begin{equation}\label{25}
  \begin{array}{ll}
&\varphi(\chi)=\pm\frac{c-b}{\sqrt{c}+\sqrt{b}\cosh(\chi)},\\[1ex]
&\xi(\chi)=2c\chi-\frac{1}{4k_1}\ln\frac{1+2w+w^2}{1-2w+w^2}
+\sqrt{c(c-b)}\ln\frac{(w+1)(\alpha_3^2+w+\alpha_3\sqrt{1+\alpha_3^2}\cosh\chi)}{(w-1)(\alpha_3^2-w+\alpha_3
\sqrt{1+\alpha_3^2}\cosh\chi)},\end{array}
\end{equation}
where $w=\alpha_3\sinh(\chi).$
The functions defined by (\ref{25}) present  two soliton
solutions shown in Fig.8 (8-1). 

{\bf 4. Two-crest soliton solutions}

Suppose $k_1>0, k_2=0, 0<b<c$ or $b\geq c>0.$ 
Then, we can solve the system (\ref{newcubic}) with
the following two-crest soliton solutions for equation (\ref{bCHFFORQ})
\begin{equation}\label{28}
  \begin{array}{ll}
&\varphi(\chi)=\pm\frac{c\sqrt{2(b+c)}\sinh(\chi)}{c+b\cosh^2(\chi)},\\[1ex]
&\xi(\chi)=-\frac12\sqrt{\frac{c}{b+c}}\chi-\ln\left(\frac{1+\sqrt{\frac{b+c}{c}}\tanh(\chi)}{1-\sqrt{\frac{b+c}{c}}\tanh(\chi)}\right).\end{array}
\end{equation}
The functions defined by  (\ref{28}) give  two-crest soliton solutions
shown in Fig.8 (8-2).

\begin{center}
\begin{tabular}{cc}
 \epsfxsize=3.5cm\epsffile{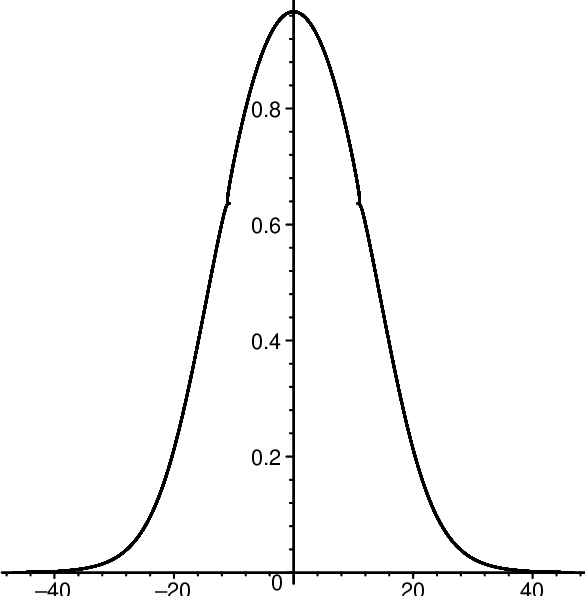} &
\epsfysize=3.5cm \epsffile{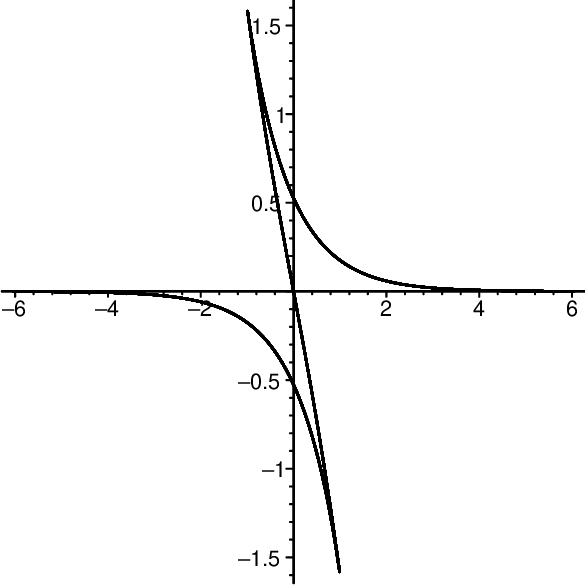}\\
(8-1) $\varphi(\xi).$ &
(8-2) $\varphi(\xi).$\\
\multicolumn{2}{c}{Fig.8   The profiles of  soliton waves with respect to
$\xi$} \\
\end{tabular}
\end{center}

{\bf Remark 4.} 
 If $B\neq 0$ in system (\ref{newcubic}), namely inhomogeneous system, then we may still
 obtain some exact smooth one-soliton solutions to equation (\ref{bCHFFORQ}) \cite{LJB2}.


\section {Conclusions and discussions}
In this paper, we have presented the Lax representation, bi-Hamiltonian structure and infinitely many conservation laws for equation (\ref{bCHFFORQ}). The peakon solutions for this equation are derived in the case $b=0$. For the one-peakon solution, we found that the complex peakon can be obtained through properly choosing the parameters $k_1$ and $k_2$ in equation (\ref{bCHFFORQ}). For the two-peakon solution, its peakon dynamical system was explicitly solved and their collisions were shown in Figures 1, 2, 3, 4. Moreover, we found that equation (\ref{bCHFFORQ}) with $k_2=0$ (namely, cubic nonlinear equation (\ref{qe})) possesses weak kink and kink-peakon interactional solutions in the case $b\neq0$. Compared with the CH equation, equation (\ref{bCHFFORQ}) has some different features, such as the dynamics of the two-peakon solutions and the existence of the {\bf weak kink and kink-peakon interactional solutions}. These differences are mainly caused by both quadratic and cubic nonlinearity in equation (\ref{bCHFFORQ}). In the paper, we also study all possible smooth one-soliton solutions. Other topics, such as cuspons, Darboux transforms, peakon stability, and algebra-geometric solutions, remain to be developed.

\section*{ACKNOWLEDGMENTS}
This work was supported 
by the U. S. Army Research Office (Contract/Grant No. W911NF-08-1-0511), the Texas Norman Hackerman Advanced Research Program (Grant No. 003599-0001-2009), 
the National Natural Science Foundation of China (Grant Nos. 11301229, 11271168, and 11171295), the Natural Science Foundation of the Jiangsu Province (Grant No. BK20130224), the Natural Science Foundation of the Jiangsu Higher Education Institutions of China (Grant No. 13KJB110009), and the China state administration of foreign experts affairs system under the affiliation of China University of Mining and Technology.

\vspace{1cm}
\small{

}

\begin{thebibliography}{99}
\bibitem{CH} R. Camassa and D.D. Holm, Phys. Rev. Lett. 71 (1993) 1661.
\bibitem{FF1} B. Fuchssteiner and A.S. Fokas, Physica D 4 (1981) 47.
\bibitem{CH2} R. Camassa, D.D. Holm and J.M. Hyman, Adv. Appl. Mech. 31 (1994) 1.
\bibitem{OR} P.J. Olver and P. Rosenau, Phys. Rev. E 53 (1996) 1900.

\bibitem{FL} A.S. Fokas and Q.M. Liu, Phys. Rev. Lett. 77 (1996) 2347.
\bibitem{DGH} H.R. Dullin, G.A. Gottwald and D.D. Holm, Phys. Rev. Lett. 87 (2001) 194501.
\bibitem{GH} F. Gesztesy and H. Holden, Rev. Mat. Iberoamericana 19 (2003) 73.
\bibitem{Q3} Z.J. Qiao, Commun. Math. Phys. 239 (2003) 309.
\bibitem{LP} P. Lorenzoni and M. Pedroni, Int. Math. Res. Not. 75 (2004) 4019.
\bibitem{LZ} Y.S. Li and J.E. Zhang, Proc. R. Soc. Lond. Ser. A 460 (2004) 2617.

\bibitem{C1} A. Constantin, V.S. Gerdjikov and R.I. Ivanov,  Inverse Problems 22 (2006) 2197.
\bibitem{C2} A. Constantin, J. Funct. Anal. 155 (1998) 352.

\bibitem{CS1} A. Constantin and W.A. Strauss, Comm. Pure Appl. Math. 53 (2000) 603.
\bibitem{CS2} A. Constantin and W.A. Strauss, J. Nonlinear Sci. 12 (2002) 415.
\bibitem{BSZ} R. Beals, D. Sattinger and  J. Szmigielski, Adv. Math. 154 (2000) 229.
\bibitem{ACH} M.S. Alber, R. Camassa, Y.N. Fedorov, D.D. Holm and J.E. Marsden, Commun. Math. Phys. 221 (2001) 197.
\bibitem{JR} R.S. Johnson, Proc. R. Soc. Lond. A 459 (2003) 1687.


\bibitem{DP1} A. Degasperis and M. Procesi, {\it Asymptotic Integrability Symmetry and Perturbation Theory eds A. Degasperis
and G. Gaeta} (Singapore: World Scientific, 1999) pp. 23-37.
\bibitem{DP2} A. Degasperis, D.D. Holm and A.N.W. Hone, Theor. Math. Phys. 133 (2002) 1463.
\bibitem{DP3} H. Lundmark and J. Szmigielski, Inverse Problems 19 ( 2003) 1241.

\bibitem{Q1} Z.J. Qiao, J. Math. Phys. 47 (2006) 112701.
\bibitem{Q11} Z.J. Qiao, J. Math. Phys. 48 (2007) 082701.
\bibitem{HW1} A.N.W. Hone and J.P. Wang, J. Phys. A: Math. Theor. 41 (2008) 372002.
 \bibitem{NV1} V. Novikov, J. Phys. A: Math. Theor. 42 (2009) 342002.
\bibitem{Fo} A.S. Fokas, Physica D 87 (1995) 145.
\bibitem{Fu} B. Fuchssteiner, Physica D 95 (1996) 229.
\bibitem{Q2} Z.J. Qiao and X.Q. Li, Theor. Math. Phys. 167 (2011) 584.
\bibitem{GLOQ} G.L. Gui, Y. Liu, P.J. Olver and C.Z. Qu, Commun. Math. Phys. 319 (2013) 731.
\bibitem{HM1} A. Himonas and D. Mantzavinos, Nonlinear Anal. 95 (2014) 499.
\bibitem{HM2} A. Himonas and D. Mantzavinos, J. Nonlinear Sci. 24 (2014) 1105.


\bibitem{GE1} J.A. Desanto, {\it Mathematical and Numerical Aspects of Wave Propagation} (SIAM, Philadelphia, 1998).
\bibitem{GE2}  A.R. Osborne, Int. Geophys. 97 (2010) 857.
\bibitem{BC} A. Bressan and A. Constantin, Arch. Ration. Mech. Anal. 183 (2007) 215.
\bibitem{LJB1} J.B. Li and H.H. Dai, {\it On the Study of Singular
Nonlinear Travelling Wave Equations: Dynamical Approach}, Science
Press, Beijing, 2007.
\bibitem{LJB2} J.B. Li, Y. Zhang, and X.H. Zhao, Int. J. Bifurcation Chaos 19 (2009) 1995.


\end{thebibliography}
\end{document}